# Controlling the stereospecific bonding motif of Au-thiolate links

*Luciano Colazzo,*[*,1,2,†] *Mohammed S. G. Mohammed,*[1,2] *Aurelio Gallardo,*[3,4] *Zakaria M. Abd El-Fattah,*[5] *José A. Pomposo,*[2,6,7] *Pavel Jelinek,*[3] *Dimas G. de Oteyza*[*,1,2,6]

[1] Donostia International Physics Center, 20018 San Sebastián, Spain.

[2] Centro de Física de Materiales (CFM-MPC), CSIC-UPV/EHU, 20018 San Sebastián, Spain.

[3] Institute of Physics, The Czech Academy of Sciences, 162 00 Prague, Czech Republic.

[4] Faculty of Mathematics and Physics, Charles University, 180 00 Prague, Czech Republic

[5] Physics Department, Faculty of Science, al-Azhar University, Nasr City E-11884 Cairo, Egypt

[6] Ikerbasque, Basque Foundation for Science, Bilbao, Spain.

[7] Departamento de Física de Materiales, Universidad del País Vasco (UPV/EHU), Apartado 1072, E-20800 San Sebastián, Spain.

**Corresponding Authors**

* luciano.colazzo88@gmail.com, * d_g_oteyza@ehu.es

**Present Addresses**

[†] Center for Quantum Nanoscience, Institute for Basic Science (IBS), Seoul 03760, Republic of Korea; Department of Physics, Ewha Womans University, Seoul 03760, Republic of Korea.

ABSTRACT.

Organosulfur compounds at the interface to noble metals have proved over the last decades to be extremely versatile systems for both fundamental and applied research. However, the anchoring of thiols to gold remained an object of controversy for long times.The RS–Au–SR linkage, in particular, is a robust bonding configuration that displays interesting properties. It is generated spontaneously at room temperature and can be used for the production of extended molecular nanostructures. In this work we explore the behavior of 1,4-Bis(4-mercaptophenyl)benzene (BMB) on the Au(111) surface, which results in the formation of 2D crystalline metal-organic assemblies stabilized by this type of Au-thiolate bonds. We show how to control the thiolate´s stereospecific bonding motif and thereby choose whether to form ordered arrays of $Au_3BMB_3$ units with embedded triangular nanopores, or linearly stacked metal-organic chains. The former turn out to be the thermodynamically favored structures and display confinement of the underneath Au(111) surface state. The electronic properties of single molecules as well as of the 2D crystalline self-assemblies have been characterized both on the metal-organic backbone and inside the associated pores.

INTRODUCTION

The production of monolayers of sulfur-containing organic compounds on noble metals was recognized readily long ago as of great interest for the development of functional interfaces.[1–7] After decades of research, the interactions between organosulfur compounds and gold have in fact become textbook examples for strongly interacting metal-organic interfaces and the investigations between organic sulfur, e.g. thiol (R-SH), and gold have shown a remarkable evolution with fruitful applications that span from biology,[8–10] drug- and medical-therapy,[11–13] to material sciences[14–18] or nanoplasmonics.[19–21]

Such a broad range of applications relies on the strong Au-S connection that occurs on the metallic surface. While the hydrogenated R-SH group would only weakly interact through coordination-type bonds with gold *via* the S lone pair electrons,[22] it is widely accepted that the Au-thiolate complex is formed after the dehydrogenation of the sulfhydryl group and the quenching of the resulting thiyl radical (RS•) with gold.[23] Small[24] or flat lying arenethioles[25] have provided excellent examples to explore the direct anchoring point of the Au-S connections by its visualization with scanning probe microscopy (SPM) and lately the use of multi-functional aryl-thiols has additionally demonstrated the possibility to use RS-Au-SR type bonding schemes to create extended and complex molecular networks on surfaces.[26,27]

In this work, combining low-temperature scanning tunneling microscopy and spectroscopy (STM LT-STM/STS), low energy electron diffraction (LEED), core level photoemission (XPS) and molecular modeling, we do not only provide a detailed picture of the whole formation process of Au-thiolates, with the associated impact on the underlying Au(111), but further show the means to control the resulting stereospecific RS-Au-SR anchoring motifs. Using 1,4-Bis(4-mercaptophenyl)benzene (BMB) we can thus choose whether to form regular arrays of triangular nanopores of $Au_3BMB_3$ with a well-defined size of ≈1.8 nm side length or wires of poly-[-S-Au-S-BMB-]$_{-n}$. The nanoporous $Au_3BMB_3$ turns out to be thermodynamically more stable on the gold surface than its linear counterpart and causes 2D quantum confinement of the Au(111) surface electrons, an interesting effect more commonly studied on surfaces of Cu(111)[28,29] and Ag(111)[30,31] than Au(111).[32]

METHODS

Clean Au(111) surfaces were prepared in situ by repeated cycles of Ar+ sputtering ($5\times10^{-6}$ mbar, 800 eV) and annealing at 680K. 1,4-Bis(4-mercaptophenyl)benzene molecules were deposited on the clean Au surface from quartz crucibles of a handmade Knudsen cell, heated at 380 K. Scanning Tunneling Microscopy (STM) micrographs were performed with a commercial Scienta-Omicron low temperature system, operating in ultrahigh vacuum (UHV) at 4.3 K. The STM tip was prepared *ex situ* by clipping a Pt/Ir wire (0.25 mm) and sharpened *in situ* by repeatedly indenting the tip few nanometers (1 to 4 nm) into the Au surface while applying bias voltages from 2V to 4V between tip and sample. In order to perform bond-resolving STM imaging, the tip apex was terminated with a CO molecule, directly picked up from the Au(111) surface, by positioning the sharp metal tip on top of it and applying a 500 ms bias pulse at -2 V (the inverted polarity of +2 V in turn controllably drops the CO back on to the surface). The imaging was then performed by measuring at constant height while applying a bias voltage within a 2.0 mV to 3.5 mV range to the tip. For spectroscopic measurements (both point spectra and conductance maps) the dI/dV signals were measured by a digital lock-in amplifier (Nanonis). STM images were analyzed by using the WSxM software.[33] The XPS analysis was performed with non-monocromatized Mg($k_\alpha$) radiation, collected by means of a SPECS Phoibos 100 hemispherical electron analyzer.

Large scale total energy Density Functional Theory (DFT) calculations of $Au_3BMB_3$ on Au(111) surface were carried out by local basis set Fireball code [34] using the general gradient approximation BLYP [35] and vdW-D3.[36] The $Au_3BMB_3$/Au(111) system was calculated with a rectangular (12×12) slab with two layers, consisting in total of 387 atoms. All atoms in the last layer were fixed. We used an optimized local basis set [37] H($R_C$(s)=5.42 a.u.), C($R_C$(s)=5.95 a.u.;

$R_C(s)$=5.95 a.u.), S($R_C(s)$=7.0 a.; $R_C(s)$=7.0 a.u.) and Au($R_C(s)$=4.50 a.u.; $R_C(s)$=5.60 a.u.; Au($R_C(d)$=4.30 a.u.). The precision of the Fireball calculations was carefully checked with the all-electron FHI-AIMS ab initio molecular simulations package[38] using the general gradient approximation PBE potential[39] and vander Waals interactions employing the Tkatchenko−Scheffler method with light basis set.[40] Both methods provide very similar optimized structures of a free-standing $Au_3BMB_3$ model triangle. Also the analysis of interaction energies of linear chains and triangles was performed using FHI-AIMS code optimizing the 2D lattice vectors. In all calculations, the Brillouin zone was sampled with Γ k-point only and the energy and force convergence were set to $10^{-5}$ eV and 0.01 eV/Å, respectively.

The bond-resolved STM images of $Au_3BMB_3$ triangles acquired with CO-tip at low bias were simulated using the probe particle SPM model.[41] The parameters of the probe particle were selected to mimic a CO-tip, using a monopole charge of -0.05 e with a lateral stiffness of 0.24 $Nm^{-1}$. The electrostatic force was included in the probe particle model calculations using the Hartree potential calculated by DFT. The dI/dV maps were calculated using the PP-STM code [42] using an s-wave probe and frontier orbitals of free-standing $Au_3BMB_3$ triangles as obtained from the Fireball code.

The electron-boundary-element-method (EBEM) was employed to extract the molecular potential landscape responsible for the confinement of Shockley state within the $Au_3BMB_3$ nanoporous network. A triangular geometry is defined for each $Au_3BMB_3$ unit, and the potentials at the boundaries and inside nanopores were set to 0.6 eV and zero, respectively [See supplementary Fig. 3 (a-b)]. The reference binding energy and effective mass of the pristine Au(111) surface state were $E_B$= 0.485 eV and $m_{eff}$ = 0.25 $m_e$.[43] The energy and spatial dependence of the local density of states (LDOSs) were obtained for single and finite island of $Au_3BMB_3$ aggregates. A complementary electron-plane-wave-expansion (EPWE) method was also used to correlate EBEM finite results with those obtained for infinite nanoporous network. The details of EBEM and EPWE methods can be found in Ref.[31,44,45]

RESULTS AND DISCUSSION

A representative STM image obtained after deposition of BMB on the Au(111) surface held at 120 K is displayed in Fig. 1a. Scattered BMB molecules and disordered molecular aggregates are observed. A closer inspection of the molecules reveals a variability of their appearance, in particular at their extremities. While three central lobes remain a common molecular trait and are depictive of a quasi-flat lying geometry of a terphenil- core, the side lobes display two types of contrast, bright or dim (see the three different examples in Fig. 1a[i],a[ii],a[iii]). Such behavior is associated with a partial dehydrogenation of the thiol groups (R-S-H) readily occurring during the condensation process. In line with a prior STM study on the thermal reactivity of benzenethiols on Au(111), the intact or undissociated thiols group are distinguished from the chemisorbed Au-thiolate (R-S-Au) by the characteristic bright and dim contrast, respectively, of the associated protrusions.[25] Thus, Fig. 1a[i], in which the terphenyl-core is sided by two bright lobes of homogeneous contrast, corresponds to the undissociated BMB molecule. Fig. 1a[ii], displays a bis-mercapto moiety with bright and dim contrast on either side, in accordance with a partial molecular dehydrogenation. Finally, the molecule in Fig. 1a[iii], sided by two dim protrusions, represents a fully dehydrogenated BMB. Noteworthy, isolated S atoms are also observed in Fig. 1a. The presence of these atoms will be further discussed below.

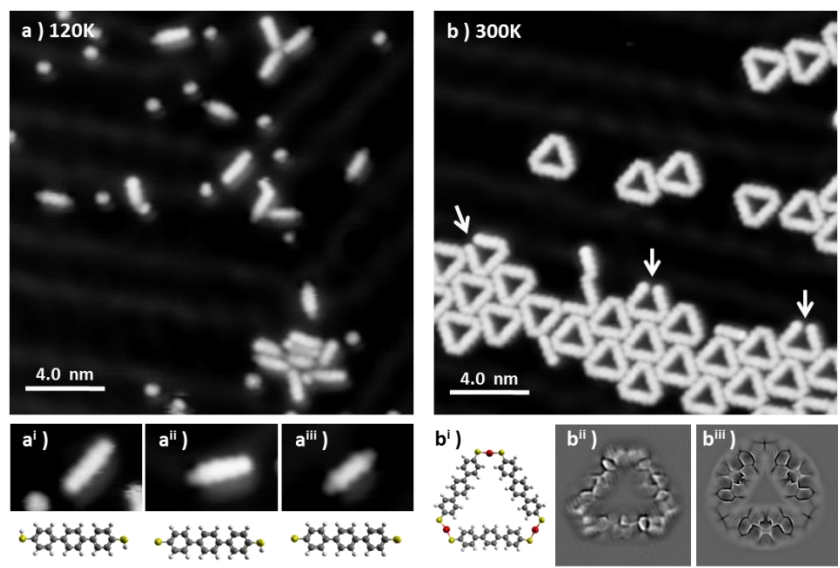

**Figure 1.** a) As-deposited, unreacted BMB molecules, 20×20 nm$^2$, V=30mV, I=10pA, T$_{depos}$=120K; a$^i$) close-up 3.0×2.0 nm$^2$ of a), and molecular model of the fully hydrogenated BMB molecule; a$^{ii}$) close-up 3.0×2.0 nm$^2$ of a), and molecular model of the partially dehydrogenated BMB molecule ;a$^{iii}$) close-up 3.0×2.0 nm$^2$ of a), and molecular model of the fully dehydrogenated BMB molecule. b) Aggregate of Au$_3$BMB$_3$ molecules obtained from a) after annealing at RT(20×20 nm$^2$, V=210mV, I=500pA), with arrows marking selected non-closed triangle structures; b$^i$) molecular model of an isolated Au$_3$BMB$_3$complex and b$^{ii}$)3.0×3.0 nm$^2$Laplace-filtered constant-height STM image with CO-functionalized probe (U = 2 mV).b$^{iii}$) Laplace-filtered image of a probe particle model simulation of a Au$_3$BMB$_3$complex.

The scenario changes completely when the sample is allowed to thermalize at room temperature (RT) and cooled again for imaging. Triangular shaped molecular units appear on the surface as isolated objects or as clusters, both lying preferentially on the fcc regions of the Au(111) surface reconstruction. The clusters aggregate with a moderate lateral ordering and appear closely packed on widened fcc regions (see Fig. 1b). As a matter of fact, the fcc/hcp reconstruction periodicity of the pristine Au(111) is lost and thoroughly modified during this process.

Within the triangular complexes the sides are discernable as terphenyl- units, while the bright connectors at the three vertexes fit with the S-Au-S bonding motif displayed in the molecular model of Fig. 1b$^\text{i}$. In order to corroborate the bonding structure of these units, constant height measurements with a CO-terminated STM tip were performed. Fig. 1b$^\text{ii}$ shows a representative Laplace-filtered tunneling-current image of an isolated triangular unit. The identification of fine molecular features results straightforward and it is possible to identify that, within each terphenyl-side, the central phenyl rings appear distorted due to the tilting imposed by steric hindrance. Importantly, each S atom of the mercapto- residue connects one Au adatom and generates a cis-type S-Au-S coordination. Here the question arises whether the Au adatom is extracted from the surface or captured from freely diffusing Au adatoms at RT. However, the fact that the average fcp/hcp period of the Au(111) herringbone reconstruction is increased or otherwise modified in this sample, implies a lower compression of the reconstructed surface layer, which in turn evidences the extraction of Au surface atoms for the thiolate-Au complex formation.

Further proof that the terphenyl-sides are linked through thiolate-Au complexes and not e.g., disulfide bonds is obtained from theoretical calculations. Both options have been simulated and relaxed by DFT (Fig. S1), revealing a formation energy of the Au$_3$BMB$_3$ complex several eV more favorable than that of the BMB$_3$ structure due to presence of covalent Au-S bonds. Besides, the Au$_3$BMB$_3$ structure shows an excellent agreement with the experimental images. Proof of it is found in the simulations with the particle probe model,[41] whose Laplace filtered image in Fig. 1b$^\text{iii}$ displays a notable similarity with the experimental data in Fig. 1b$^\text{ii}$. Particularly revealing is the presence of characteristic sharp edges at the presumable thiolate bonds of the $Au_3BMB_3$ structure due to lateral relaxation of CO-tip, which is completely missing in the case of the $BMB_3$ structure. From the total energy DFT simulations we can also deduce that the formation of strong covalent bonds between gold and sulfur atoms introduces a slight lateral distortion of the relatively soft BMB molecular units. These deformations are clearly visible in the high-resolution STM images acquired with CO-tip.

It is also worth pointing out that, occasionally, incomplete triangles are found (see arrows in Fig. 1b). A closer look at the loose ends of these metal organic complexes evidences only three phenyl rings and the lack of a thiol/thiolate group, presumably associated with a thermal degradation occurring during the sublimation of the BMB powder. Indeed, following the deposition at 120K, it was found that a large amount of elemental sulfur adsorbs on the Au surface. In addition to single atoms or few-atom clusters, numerous 2D sulfur islands are also generated (Fig. S2a). Combining STM and LEED on this sample (Fig. S2c and S2d) it was found that S adsorbs in the well-known $(\sqrt{3}\times\sqrt{3})R30$ phase,[46,47] not affecting the herringbone reconstruction of the underlying Au (Fig. S2b). This phase is known to display limited stability, which may be the reason why we are only able to observe it at 4.3K after its condensation on the Au surface held at 120K. Finally, by letting the system thermalize to RT and cooling again to 4.3 K for imaging, a complete desorption of the atomic sulfur and a transformation of the scattered single molecules into the triangular metal-organic complexes is observed.

Considering the stereochemistry of the molecular product obtained in the sub-monolayer regime, and in particular focusing on the S-Au-S connectors of the molecular structure, it is remarkable that the cis- configuration of the system, which allows the formation of the triangle, occurs with such high specificity. We have explored the possibility to stabilize the trans-conformation of the S-Au-S connections in order to generate linear chains of poly-bis-mercaptobenzene. To do so, the molecular deposition of the surface of Au has been performed with a high molecular flux (by increasing the deposition rate to 300 monolayers/h), low temperature (the Au(111) surface being held at 120 K) and full monolayer (ML) coverage. As can be observed from the bright termini of the BMB molecules (Fig. 2a), under these conditions the –SH groups remain intact and the Au(111) reconstruction underneath the molecular layer is not affected (Fig. S3). The thiol´s bright-end contrast disappears when the system is annealed to RT. Although in this temperature range the transformation of the overlayer in linear Au-thiolate chains does not occur quantitatively, the molecular arrangement loses its original periodicity and concurrently the surface morphology, i.e. the herringbone, starts being affected, as mentioned above, due to the cumulative extraction of surface atoms by the thiolates (Fig. S3). When the system is brought to 400 K a full transformation into metal-organic chains is obtained and the Au(111) herringbone reconstruction is not observed anymore. The molecular layer appears as a series of long chains clearly bridged by a trans- conformation of the S-Au-S connectors (Fig. 2b). The chains extend over several tens of nanometers although their lateral order remains rather limited. Finally, the metastability of the trans- organometallic chain is revealed when the system is annealed at 500K. After the annealing, the system switches to the triangular nanopore array (Fig. 2c). The lower density of this phase readily denotes a substantial molecular desorption. Indeed, annealing above 500 K the molecular overlayer undergoes degradation and desorbs almost entirely. These findings imply that the triangular complexes are more stable than their linear counterparts. To corroborate this hypothesis, we carried out total energy DFT simulations of linear and triangle structures and compared the formation energies of their 2D assemblies. First, we compare the formation energies of a single 1D chain and triangle (see Fig. S4), which indicates that the individual chains are energetically slightly favorable by 18 meV/molecule. However, the situation changes when 2D assemblies are formed. After optimizing the lattice parameters for the free-standing assemblies, the binding energy per molecule turns out to be 230 meV larger for the triangle complexes. This indicates that it is the cumulative dispersion forces involved in the formation of the 2D molecular arrays which render the triangular aggregates substantially more stable than the chain-type aggregates.

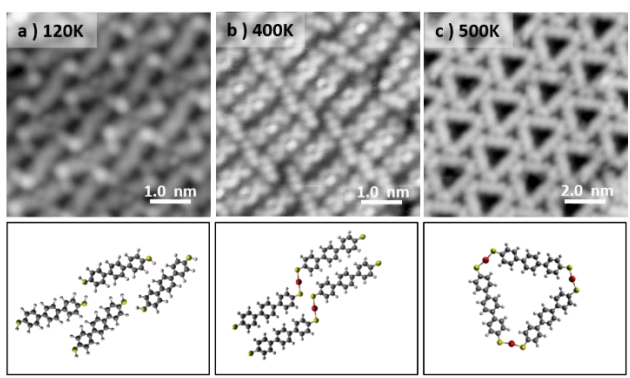

**Figure 2.** Formation of the linear poly-[-S-Au-S-BMB-]$_n$ at RT, in the high coverage regime and subsequent trans-to-cis isomerization into Au$_3$BMB$_3$ at higher temperatures on Au(111). a) As deposited($T_{depos}$=120K), unreacted BMB molecules(5.0×5.0 nm$^2$,V=50mV, I=80pA). b) Poly-[-S-Au-S-BMB-]$_n$ obtained from a) after termalization at RT (5.0×5.0 nm$^2$, V=-100mV, I=300pA).c) Au$_3$BMB$_3$ obtained after annealing of b) at 500K (10.0×10.0 nm$^2$, V=-500mV, I=10pA).

The hypothesis that in the high coverage regime, the adsorption of BMB molecules on Au(111) at 120K occurs with intact thiol functionalities and that at higher temperature the chemical environment of S atoms changes, is also corroborated by XPS measurements. The as-deposited system at 120K shows the S 2p3/2 peak at 163.3 eV (Fig. S3a) characteristic of physisorbed or unbound thiol groups,[48–50] while the sample annealed to RT shows a shift to higher binding energies, namely 162.4 eV (Fig. S3b). This value is in agreement with several studies with different precursors regarding the RT formation of the Au-thiolate.[27,50,51] By further increasing the annealing temperature to 450K the XPS spectrum remains nearly unchanged (Fig. S3c). At this temperature, the chemisorbed sulfurs no longer coordinate to surface atoms but to extracted adatoms. Besides, 450 K also represents the onset of the formation of the triangular phase. Therefore, once the covalent S-Au bond is formed, the Au adatom ablation or the conformational trans/cis changes seem to render virtually identical signal with XPS, nonetheless characteristic of the RS−Au−SR bonding motif.

In order to describe the 2D crystal lattice periodicity and to extract the adsorbate unit cell, LEED analysis was performed on the thermodynamically stable network of triangular complexes. It is worth pointing out that the adsorption of the triangles leads to the formation of a total of 6 domains that can be considered as the sum of two sets of 3 symmetry-equivalent sub-domains. The triangular units adsorb on the Au(111) by adopting two configuration, namely "up" and "down" as depicted in Fig. 3a and both phases are observed to segregate during the formation of the close packed molecular arrays. The two sub-domains are related to one another being their own specular image and on a local level spontaneous symmetry breaking occurs (see unit cell representations and the boundary region highlighted with the yellow zig-zag line in Fig. 3a). For these reasons the LEED pattern shows as a superposition of the enantiomorphous lattices since they are probed simultaneously during the analysis. We have simulated the diffraction pattern for a superlattice characterized by the epitaxial matrix (8, 1 / -1, 7), associated with a hexagonal unit cell with lattice vectors a=2.17 nm and a 6.59º rotation with respect to the underlying substrate lattice, all in agreement with our STM images. In Fig. 3b and 3c the diffraction patterns have been simulated for each up and down domain. Finally, when the two images are superimposed, an excellent fit with the experimental LEED pattern is obtained (Fig. 3d), confirming the commensurate epitaxy. The stabilization energy of the self-assembled long-range-ordered 2D nanostructures arises from the intermolecular interactions, as already evidenced in the calculations of Fig. 2d, while a favorable interplay of molecule-substrate interactions ensures the commensurability of this molecular layer with the underlying Au(111) surface.

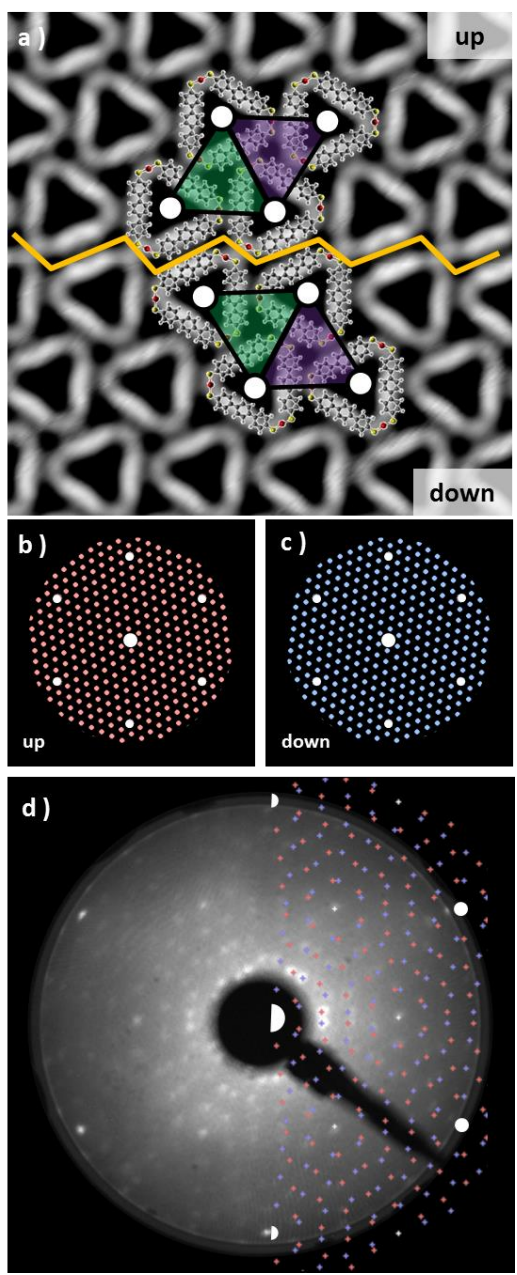

**Figure 3.** STM topography of a domain symmetry breaking of Au$_3$BMB$_3$ , 10×10 nm$^2$, V=260mV, I=400pA. The yellow zig-zag line indicate the symmetry breaking region. Molecular models are superimposed to the STM topography and their mirror symmetry relation is indicated with a colored unit cell; b) and c) Simulation of LEED patterns for the up and don domains respectively; d) experimental LEED pattern at 51 eV with superimposed LEED pattern simulation.

We have also probed the electronic properties of the Au$_3$BMB$_3$ complexes by STS and theoretical calculations, both on their metal-organic backbone, as well as inside the pores. Regarding the former, Fig. 4b depicts detailed density of states (DOS) calculations of fully optimized Au$_3$BMB$_3$ structures in a free-standing configuration (top) as well as adsorbed on Au(111) (bottom). The DOS of the free-standing complex displays well-defined orbitals, whose character is pictured in Fig. 4a for the frontier states. Upon absorption on Au(111), the DOS appears broadened (Fig. 4b, lower panel) and the atomic orbitals of the complex´ Au atoms hybridize strongly with surface atoms, causing an overall energy shift. However, the frontier orbitals are preserved and the same orbital character is now distinguished at lower energies (marked with arrows in Fig. 4b, bottom). The calculated DOS indeed shows good agreement with our experimental data. Probing with STS on the terphenyl-units of the arms (black curve) we identify two clear resonances at -1.45 eV and +2.28 eV (Fig. 4e). Except for minor differences in energy due to size effects and the associated electron confinement,[52] the energies and local density of states distribution (Fig. 4d) very much resemble previous data on poly-paraphenylene on Au(111).[53] In contrast, spectra measured on the Au-thiolate connectors display a rather featureless signal, although evidencing a notable density of states within the previously mentioned terphenyl gap, including a shoulder around -0.7 eV (Fig. 4e). The corresponding conductance maps at -1.45 eV, -0.7 eV and +2.28 eV are displayed in Fig. 4d. As mentioned above, while the maps at -1.45 eV and +2.28 eV resemble poly-paraphenylene orbitals,[53] the data at -0.7 eV reveal a spatially structured electronic density of states with strongest weight around the Au-thiolate connection region. Notably, the three maps display remarkable agreement with simulated dI/dV images (Fig. 4c) of the three frontier orbitals pictured in Fig. 4a, providing a fully coherent picture of theory and experiment.

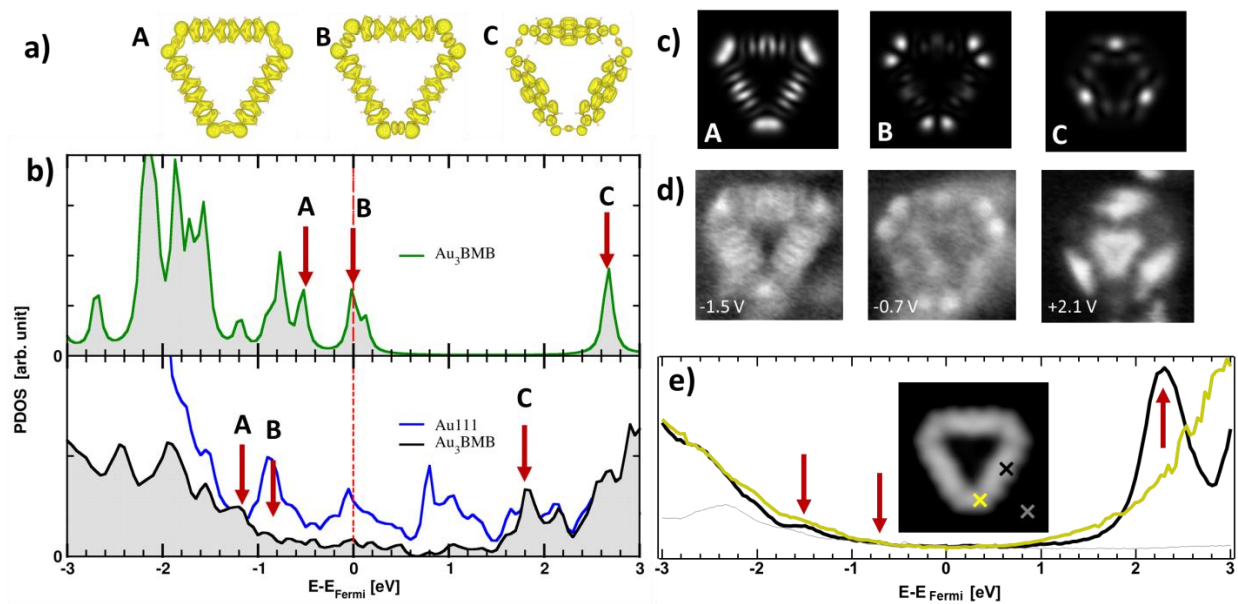

**Figure 4.** a) Simulated frontier orbitals A, B, and C of an isolated $Au_3BMB_3$ complex. B) Calculated density of states of a free-standing (top) and adsorbed (bottom) $Au_3BMB_3$ complex, marking with arrows the energies associated with A, B and C orbitals. c) Simulated dI/dV maps of A, B and C. d) Experimental dI/dV maps measured at -1.5 eV, -0.7 eV and +2.1 eV. e) Experimental dI/dV point spectra (Vrms=15mv at 731Hz, closing feedback parameters V=1V, I=200pA) measured on the terphenyl arm (black), Au-thiolate region (yellow), and reference spectrum on the substrate (grey), as marked in the inset by the colored crosses superimposed to the STM topography of an isolated $Au_3BMB_3$ on Au(111) (3.5×3.5 $nm^2$, V=50mV, I=100pA).

We now focus on the STS spectra recorded inside the pores. Interestingly, they reveal a substantially different signal from that detected on the bare Au(111) surface, in all cases evidencing the disappearance of the sharp Au(111) surface state onset at ≈ -0.48 eV (grey spectrum in Fig. 5c), and the concomitant appearance of a broad signal at higher energies. Figure 5c displays a comparative series of representative dI/dV spectra within isolated triangular units and differently positioned units within a 2D island. The signal within isolated triangles is a broad band centered around +200 mV (red curve), which we assign, in line with previous reports on porous networks,[28–32,54] to electronic states associated with the confinement of the surface state electrons (n=1 resonance). That is, the adsorbed $Au_3BMB_3$ complex acts as a scattering potential for the surface state electrons, which are thus confined into the nanopores and as a result upshifted in energy. However, there is also certain transmission across such scattering barrier, which partially accounts to the notable resonance width. Interestingly, when assembled into ordered arrays, those electronic states leaking through the confining barrier can couple with the states of neighboring pores and end up forming well-defined bands.[28,32,54] This effect can be nicely observed comparing the previous spectrum with the spectra on regular complex arrays. The triangular complex at the island edge is sided by other complexes at two of its sides and leaves only one side leaking electronic density of states toward the bare Au(111) surface. As a result, the pore´s density of states readily appears much better defined than for the isolated triangle, but still utterly different from that of the following complex toward the island interior, surrounded by neighbors on its three sides. For such "bulk" complexes a much sharper and clearly structured density of states appears in between about -200 meV and 230 meV (displaying a strong maximum at 205 meV). As mentioned above, the electronic states at discrete energies within the pores, couple giving rise to bonding and anti-bonding states. For extended arrays, these form a continuous band whose bandwidth (limited by bonding and antibonding states at the low and high energy side, respectively) is proportional to the interaction between pores.

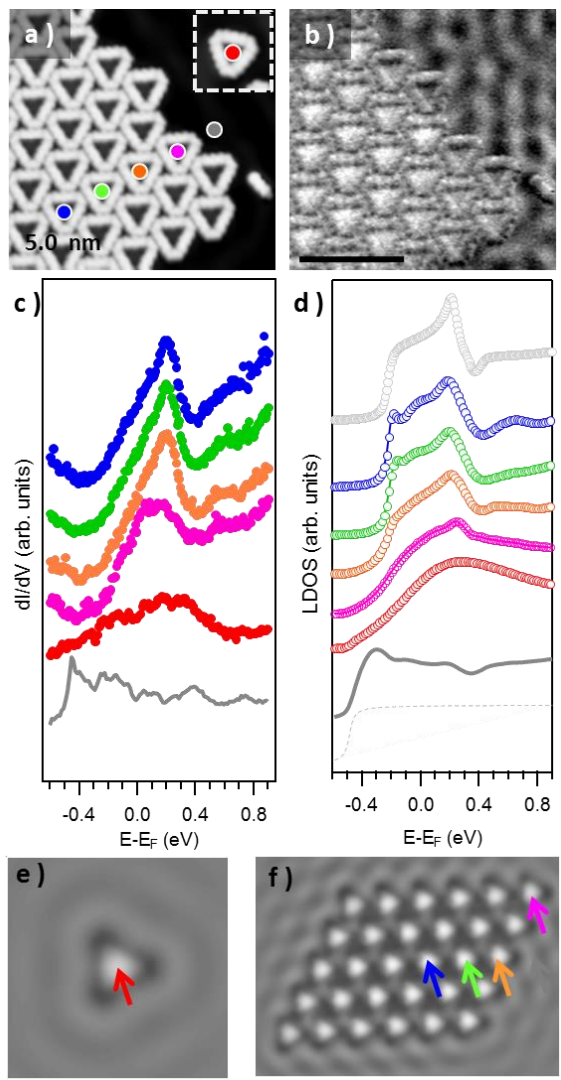

**Figure 5.** a) STM topography of an array of condensed triangular Au$_3$BMB$_3$ complexes (a single complex is displayed in the inset) obtained after annealing at 450K (U =210mV, I=400pA) and b) corresponding conductance map recorded simultaneously (U$_{ac}$ = 10mV at 731Hz). c) STS point spectra recorded inside the pores of the single triangle and of the triangles forming a condensed island, at the locations marked with the corresponding colored dots in panel (a) (U$_{ac}$=15mV at 731Hz). d) Simulated LDOS spectra at the pore center of an isolated triangle (red), of triangles displaying one (orange), two (green), and three (blue, violet) neighbors within finite molecular island, as well as for an infinite network (light grey). The LDOS close to an island and for the pristine Au(111) substrate are shown in dark solid and light dashed grey, respectively. Spatially resolved LDOS map taken at the resonance energy for (e) an isolated triangle and (f) a molecular island.

In order to quantify the degree of inter-pore coupling, i.e. the strength of the confining potential, we performed EBEM simulations for an isolated triangle and for a finite molecular array (Fig. S5). In first approximation, we discard effective mass renormalization ($m_{eff}$ = 0.25 $m_e$), and use the band minimum of the Shockley surface state on pristine Au(111) as the energy reference ($E_B$= -0.485 eV). For a scattering potential of 0.6 eV, the onset of the surface state is upward shifted to -0.2 eV and the experimental resonance located at ≈0.2 eV is reproduced, Fig. 5(d). The LDOS on the substrate close to the molecular island is only slightly modified (dashed grey) with an onset that nearly coincides with that of the pristine Au(111) surface state (solid grey). The evolution/narrowing of the resonance width by going from an isolated triangle (red), to a complex at an island edge displaying one (orange) or two (violet) neighbors, and eventually to the interior of the network (green and blue) agrees remarkably well with the experiment. Indeed, even for such a small island, the LDOS at the network's interior largely resembles the calculated one for an infinite network as obtained from EPWE calculations (black curve). The spatial distribution of the n = 1 pore state at the resonance energy for both a single triangle and a finite molecular island are presented in 2D-LDOS maps in Fig. 5e and Fig. 5f, respectively. The good matching between the LDOS obtained from EBEM and EPWE calculations allows the estimation of an 80 meV zone-boundary gap (Fig. S5d).

CONCLUSIONS

In conclusion, we report the controlled formation of stereospecific RS-Au-SR bonding motifs allowing for the selective stabilization of linear or porous Au-thiolate isomers. The investigation of the formation mechanism of covalent S-Au bonds starting from physisorbed thiols has revealed that the selective stabilization of the thiolate isomers is strictly dependent on the surface coverage and temperature. The linear isomers are thermodynamically less stable than the porous counterpart, but can be stabilized kinetically. An analysis of the electronic properties of the more stable porous arrays of $Au_3BMB_3$ has revealed that, besides the electronic states associated to the metal-organic backbone, the pores act as confinement barriers to the surface state electrons of the enclosed Au(111) patch. We can thus obtain 0-dimentional quantum dot states from single $Au_3BMB_3$ molecules, which can further couple into well-defined bands within extended ordered arrays of the Au-thiolate complexes. Interestingly, the finite size effects at the array´s edges are unambiguously observed.

CONFLICTS OF INTEREST

There are no conflicts to declare

ACKNOWLEDGMENT

The project leading to this publication has received funding from the European Research Council (ERC) under the European Union's Horizon 2020 research and innovation programme (grant agreement No 635919), and from the Spanish Ministry of Economy, Industry and Competitiveness (MINECO, Grant No. MAT2016-78293-C6-1-R). P.J. and A.C. acknowledge support from Praemium Academie of the Academy of Science of the Czech Republic, MEYS LM2015087 and GACR 18-09914S and Operational Programme Research, Development and Education financed by European Structural and Investment Funds and the Czech Ministry of Education, Youth and Sports (Project No. CZ.02.1.01/0.0/0.0/16_019/0000754).

# Supplementary Information

# Controlling the stereospecific bonding motif of Au-thiolate links

*Luciano Colazzo,*[*,1,2,†] *Mohammed S. G. Mohammed,*[1,2] *Aurelio Gallardo,*[3] *Zakaria M. Abd El-Fattah,*[4] *José A. Pomposo,*[2,5,6] *Pavel Jelinek,*[3] *Dimas G. de Oteyza*[*,1,2,5]

[1]Donostia International Physics Center, 20018 San Sebastián, Spain.

[2] Centro de Física de Materiales (CFM-MPC), CSIC-UPV/EHU, 20018 San Sebastián, Spain.

[3]Institute of Physics, The Czech Academy of Sciences, 162 00 Prague, Czech Republic.

[4] Physics Department, Faculty of Science, al-Azhar University, Nasr City E-11884 Cairo, Egypt

[5]Ikerbasque, Basque Foundation for Science, Bilbao, Spain.

[6] Departamento de Física de Materiales, Universidad del País Vasco (UPV/EHU), Apartado 1072, E-20800 San Sebastián, Spain.

**Calculations of different BMB trimer structures**

The bonding between the BMB molecules has been studied theoretically by means of DFT calculations. Two structures, with the molecules connected through thiolate-Au and disulfide bonds, were considered. The relaxed structures are shown in figure S1. The formation of the structures connected by S-Au-S is found to be arround 5 eV more favorable than the S-S linked one. The interaction between molecules in absence of adatoms is hard to attain, since the molecules tend to stay linked mainly with the substrate. This is reflected in the relaxed structure of three BMB molecules put close each other trying to form a BMB3 compound by S-S linkers (Figure S1b). In this scenario the molecules repels each other and links the S atoms to the substrate, with a S-S distance of 4.77 Å in average, while the distance between the sulfur and substrate atoms is 2.58 Å in average. In addition, as readily commented in the main text, the structure with S-Au-S links agrees much better with the structure observed experimentally.

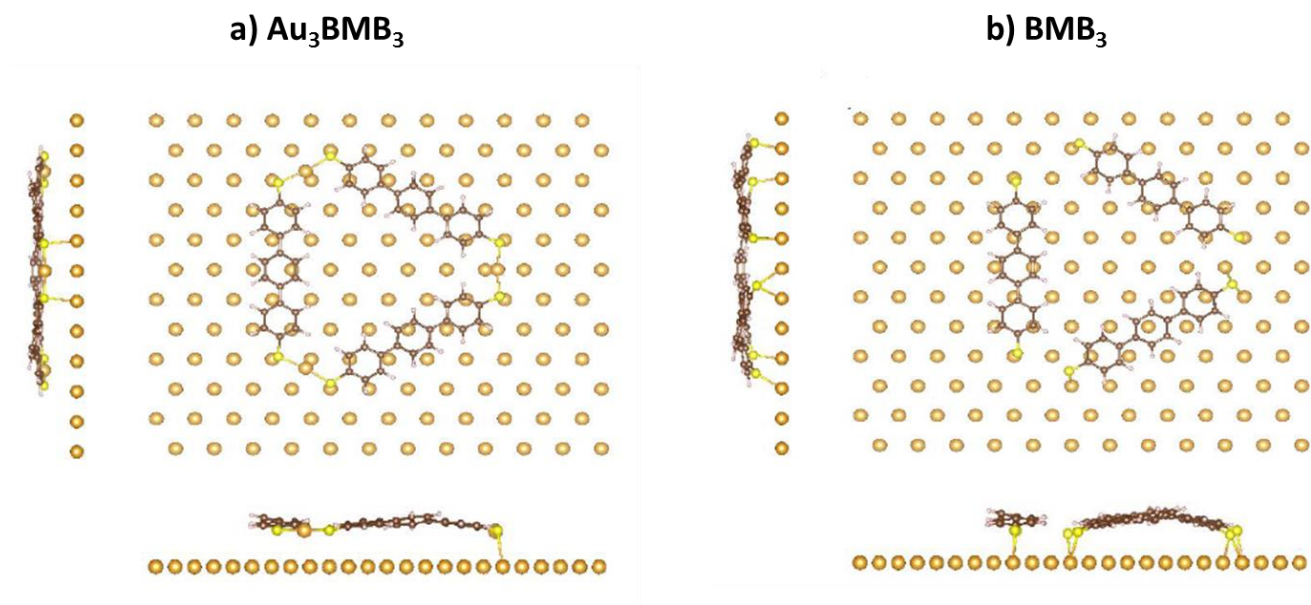

**Fig. S1. Top and lateral views of (a) the Au$_3$BMB$_3$ and (b) the BMB$_3$ relaxed structures.**

**BMB 1.7 ML/h deposition on Au(111) held at 120K**

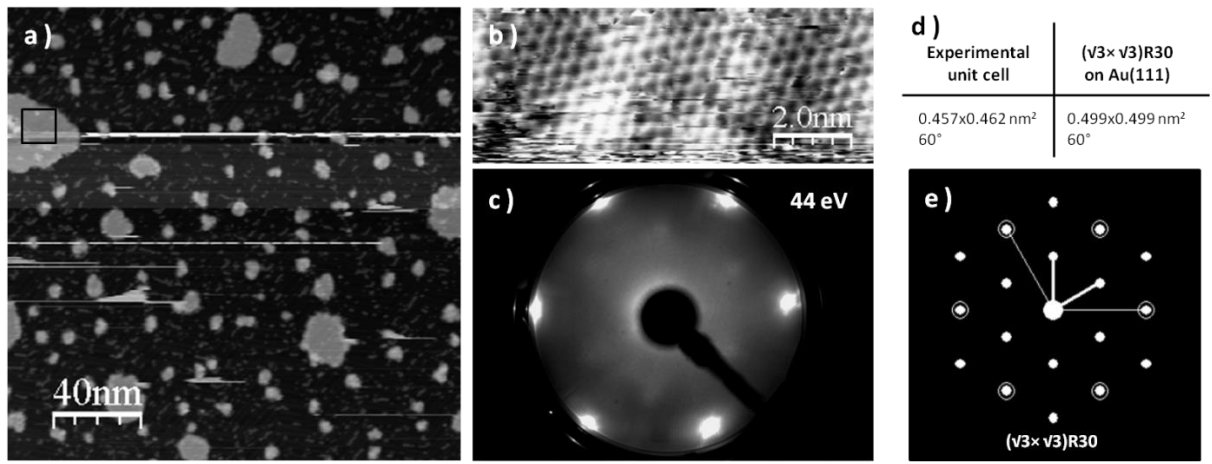

Fig. S2. STM topography after the deposition of BMB molecules at 120K, 200×200 nm², V=600mV, I=25pA. The black square on the left side of the image corresponds to b) STM topography of a S island 10×4 nm², V=-50mV, I=40pA c) LEED patterns; d) comparison chart for the experimental S unit cell and the simulated ($\sqrt{3}\times \sqrt{3}$)R30 unit cell on Au(111) (substrate unit vector 0.288 nm) e) simulated LEED pattern of the ($\sqrt{3}\times \sqrt{3}$)R30 unit cell on Au(111).

After the adsorption of BMB molecules on Au(111) at 120K and immediately cooled down to 4.3K it was possible to observe that an abnormal quantity of atomic sulfur coadsorbs on the surface (Figure S1a). The ratio BMB molecules to S is 1:98. The extra sulfur is presumably due to thermal degradation processes that occur in the crucible during the sublimation, indeed in this sample numerous BBM molecules are observed to adsorb without thiol groups. The latter are responsible for the formation of incomplete trangles as discussed in the main text. The sulfur islands do not affect the herringbone reconstruction of the Au surface (Figure S2b) and adopt a close packed hexagonal motif. By performing LEED analysis on this system (Figure S2c) it is possible to identify a ($\sqrt{3}\times \sqrt{3}$)R30 aggregation pattern on the surface of Au(111). The unit cell parameters extracted experimentally from the STM topographies (Figure S2d) display unit vectors 0.457x0.462 nm² and an angle of 60° between them. In good agreement with the simulation of a ($\sqrt{3}\times \sqrt{3}$)R30 obtained with the LeedPat software. In Figure S2e the simulation of the LEED patter is reported.

**XPS (Mg Kα) on DTTP/Au(111)**

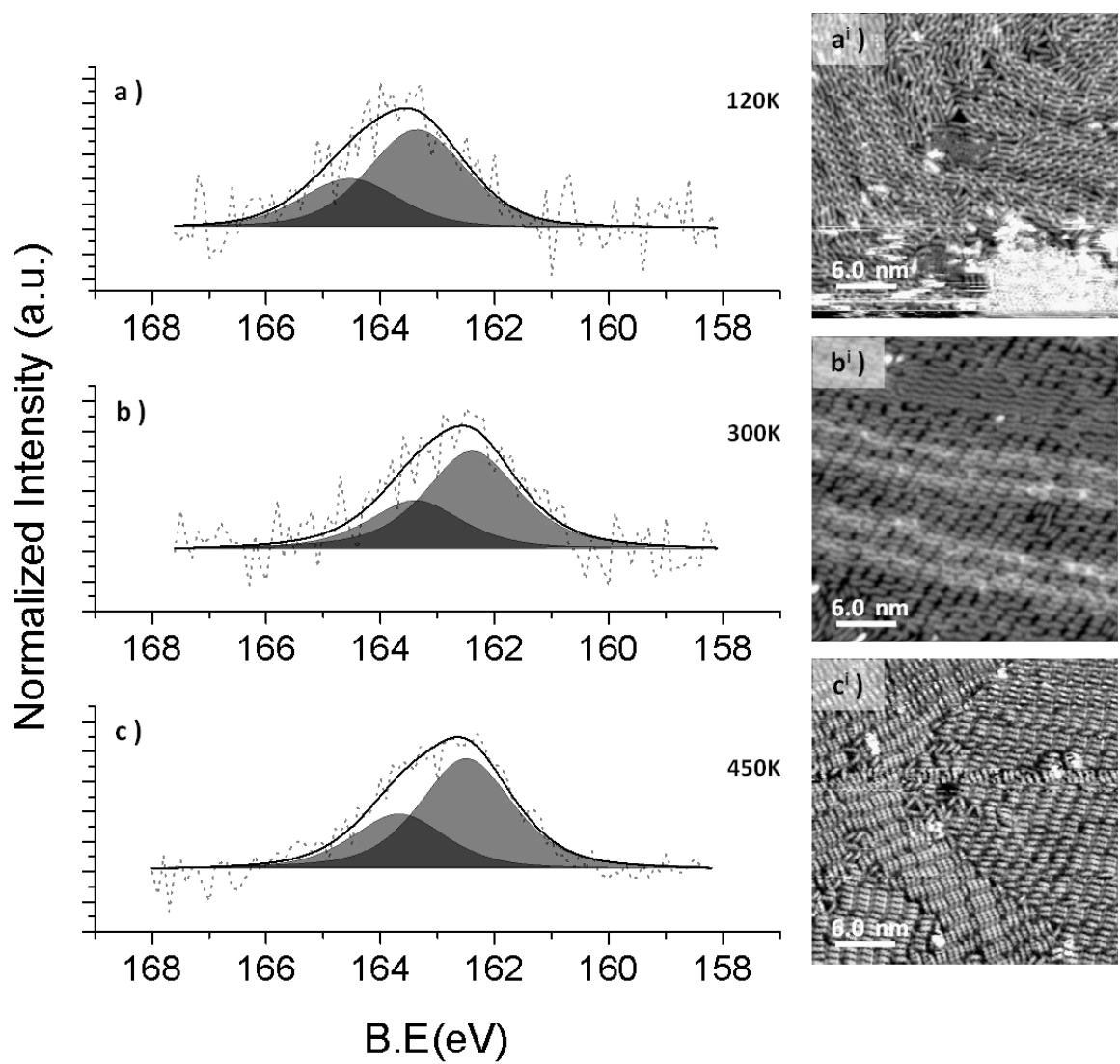

Fig. S3. Mgkα XPS spectra of S2p acquired a)after deposition at 120K b) after RT thermalizaiton and c) after annealing at 450K. The annealing steps were carried out on the same sample. S 2p 1/2 and 3/2 doublet fitted with a fixed energy separation of 1.18 eV. ai) STM topography 30×30 nm$^2$, V=-50mV, I=10pA for the system deposited at 120K and recorded at 4.3K bi) STM topography 30×30 nm$^2$, V=-50mV, I=80pA for the system thermalized at RT and recorded at 4.3K and ci) STM topography 30×30 nm$^2$, V=-500mV, I=100pA for the system annealed at 450K and recorded at 4.3K

The XPS analysis on the high molecular density sample, performed at different temperatures, reveal important changes relative to the S2p XPS signal. The as-deposited system at 120K reveals a S2p signal where the S 2p3/2 peak at 163.3 eV (Figure S3a) characteristic of physisorbed, or unbound thiol groups. In figure S3ai a representative STM topography is reported where it is possible to observe the unaffected herringbone pattern of the Au(111) surface. The sample annealed at RT shows the S 2p3/2 peak locates at 162.4 eV (Figure S3b). A representative image for this system is displayed in Figure S3bi where It is possible to observe the onset of the deformation of the Au(111) herringbone reconstruction. The molecules now appear with dimmer termini following the chemisorptions with the Au surface atoms. Occasionally, in this system, also wires of linear metal-organic Au-thiolates are found. Since the latter are formed by ablation of the Au-surface atoms the surface regions underneath do not display the herringbone reconstruction anymore. By further increasing the annealing temperature to 450K the XPS spectrum remains nearly unchanged (Figure S3c) and in the corresponding STM topography (Figure S3ci) it is possible to observe that the linear metal-organic complex are the dominant product. At this temperature it is also observed the onset of the formation of the triangular complexes and the herringbone reconstruction of the Au(111) is completely lost. By annealing over 500K the XPS spectra become heavily attenuated due to the substantial desorption of the molecular overlayer. Besides, the onset of side reactions (formation of thioethers)[1] and the broadening of the spectra makes the identification of the sulfur species via XPS analysis impossible.

---

[1] Rastgoo-Lahrood, A.; Martsinovich, N.; Lischka, M.; Eichhorn, J.; Szabelski, P.; Nieckarz, D.; Strunskus, T.; Das, K.; Schmittel, M.; Heckl, W. M.; et al. From Au–Thiolate Chains to ThioetherSierpiński Triangles: The Versatile Surface Chemistry of 1,3,5-Tris(4-Mercaptophenyl)Benzene on Au(111). *ACS Nano* **2016**, *10* (12), 10901–10911. https://doi.org/10.1021/acsnano.6b05470

**Comparison of formation energies per molecule for different structures.**

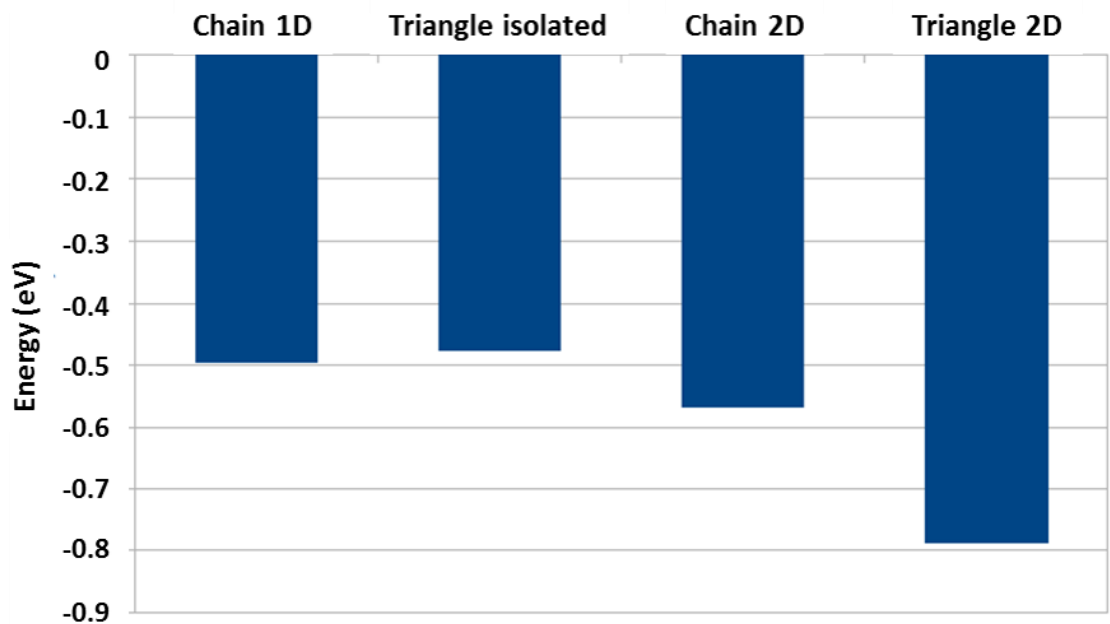

**Fig. S4.** Formation energies per molecule calculated by DFT for linear and triangular complexes as single entities, as well as in 2D assemblies.

**Geometries used in EBEM calculations**

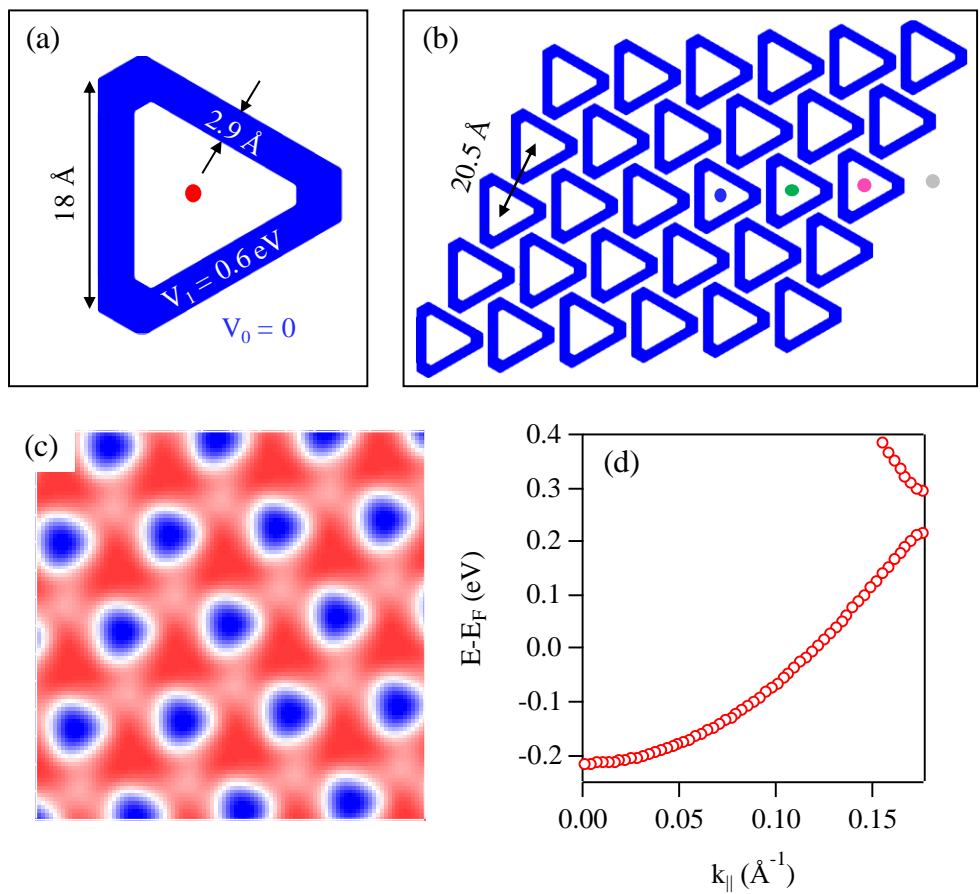

Fig. S5. The geometries used for EBEM calculations for the (a) Single triangle and (b) a 5x6 molecular island. The white background stands for the Au(111) substrate, while the blue segments define the molecular backbone with a scattering potential of 0.6 eV. The molecular size and periodicity are indicated on the figure. The circles in (b) mark the location at which the LDOSs presented in Fig. 4 is taken. (c) The 2D-LDOS for an infinite network taken at the resonance energy ( E = 0.205 eV) (d) The EPWE-calculated surface state dispersion for an infinite triangle network, showing a 80 meV gap opening at the zone boundary.